\documentclass[12pt]{article}
\usepackage{geometry}
\usepackage{amsmath, amssymb, graphicx,fullpage,color,mathtools,amsthm,xcolor}
\usepackage{caption, subcaption}
\usepackage{mathtools}  

\usepackage[numbers]{natbib}

\usepackage[normalem]{ulem}

\usepackage{microtype}
\usepackage{hyperref,color}
\usepackage{dirtytalk}
\usepackage{fix-cm}
\usepackage[linesnumbered, ruled,vlined]{algorithm2e}

\definecolor{webgreen}{rgb}{0,.35,0}
\definecolor{webbrown}{rgb}{.6,0,0}
\definecolor{RoyalBlue}{rgb}{0,0,0.9}
\definecolor{purp}{rgb}{0.6,0.05,0.8}
\definecolor{ora}{rgb}{0.7,0.35,0.02}

\hypersetup{
   colorlinks=true, linktocpage=true, 
   urlcolor=webbrown, linkcolor=RoyalBlue, citecolor=black,
   pdfauthor={Gary P. T. Choi, Mahmoud Shaqfa},
   pdfsubject={Hemispheroidal parameterization and harmonics}
}

\usepackage{todonotes}

\begin{document}

\author{Mahmoud Shaqfa\href{https://orcid.org/0000-0002-0136-2391}{\protect\includegraphics[scale=.050]{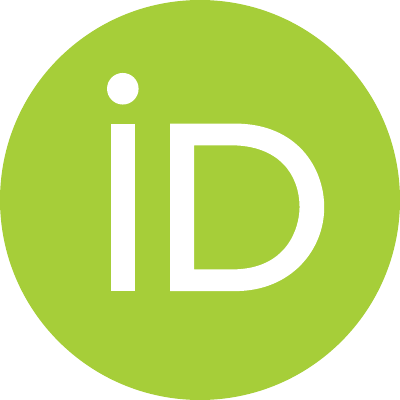}}$^{1,2,*}$\\
\\
\footnotesize{$^{1}$Croatian Centre for Earthquake Engineering, University of Zagreb, Zagreb, Croatia}\\
\footnotesize{$^{2}$Department of Mechanical Engineering, Massachusetts Institute of Technology, Cambridge, MA, USA}\\
\footnotesize{$^\ast$To whom correspondence should be addressed; E-mail: mshaqfa@mit.edu}
}

\title{Morphology-Preserving Remeshing Approach to Particulate Microstructures via Harmonic Decomposition}
\date{}
\maketitle

\begin{abstract}
Harmonic decomposition of surfaces, such as spherical and spheroidal harmonics, is used to analyze morphology, reconstruct, and generate surface inclusions of particulate microstructures. However, obtaining high-quality meshes of engineering microstructures using these approaches remains an open question. In harmonic approaches, we usually reconstruct surfaces by evaluating the harmonic bases on equidistantly sampled simplicial complexes of the base domains (e.g., triangular spheroids and disks). However, this traditional sampling does not account for local changes in the Jacobian of the basis functions, resulting in nonuniform discretization after reconstruction or generation. As it impacts the accuracy and time step, high-quality discretization of microstructures is crucial for efficient numerical simulations (e.g., finite element and discrete element methods). To circumvent this issue, we propose an efficient hierarchical diffusion-based approach for resampling the surface---i.e., performing a reparameterization---to yield an equalized mesh triangulation. Analogous to heat problems, we use nonlinear diffusion to resample the curvilinear coordinates of the analysis domain, thereby enlarging small triangles at the expense of large triangles on surfaces. We tested isotropic and anisotropic diffusion schemes on the recent spheroidal and hemispheroidal harmonics methods. The results show a substantial improvement in the quality metrics for surface triangulation. Unlike traditional surface reconstruction and meshing techniques, this approach preserves surface morphology, along with the areas and volumes of surfaces. We discuss the results and the associated computational costs for large 2D and 3D microstructures, such as digital twins of concrete and stone masonry, and their future applications.
\end{abstract}


\section{Introduction}
\label{sec:intro}
\par
Studying the morphology of surfaces and composite microstructures is crucial for applications in engineering mechanics, biomedical imaging, and computer graphics. Nowadays, harmonic decomposition approaches are widely used for studying the morphology of arbitrary surfaces, with a focus on genus-0 open and closed surfaces. These methods are used to study surfaces generated by scanning technologies and are typically represented as point clouds or triangulated surfaces. The discretization quality of such scanning techniques is often insufficient for numerical simulations and requires intensive preprocessing. Furthermore, the reconstructed details on surfaces are not uniformly represented as they are oversampled in some areas and undersampled in others. Powerful surface remeshing pipelines already exist, such as \cite{Hoppe1993, Remacle2010, Kazhdan2013}. However, to the best of the author's knowledge, the literature is scarce on high-quality morphology-preserving remeshing approaches that leverage harmonic decomposition without relying on surface oversampling or oversimplification.
\par
To study the morphology of 3D closed surfaces, Brechb{\"u}hler et al. (1995) \cite{Brechbuhler_1995} proposed a spherical parameterization (mapping) approach to map genus-0 surfaces onto a sphere that is followed by decomposing the surface data using the spherical harmonics (SH). The SH approach is widely used in analyzing and generating particle-based heterogeneous microstructure \cite{Qian_2012_thesis, Zhou2016, Xiong2021, Wang2021, SHAQFA2022, Zheng2022, Huang2023, Zhao2023, Hu2024}. However, due to the large mapping errors on spheres, the traditional SH approach suffers from spurious noise that appears in the reconstruction stage. To alleviate this effect, Shaqfa and van Rees (2024) \cite{Shaqfa2024_SOH} proposed the spheroidal harmonics (SOH) approach as a generalization of the traditional SH. The SOH approach proposed a simple and computationally efficient mapping methods that allow for mapping arbitrary genus-0 surfaces into oblate or prolate domains. These approaches have multiple mechanical applications. For instance, Grigoriu et al. (2006) \cite{Grigoriu2006} used SH to generate statistically similar particles, alongside studying the mechanical behavior of particles via the discrete element method (DEM) as in \cite{Imaran2024, Capozza2021}.
\par
For 3D open surfaces, multiple papers have been recently proposed to study genus-0 and single-edged objects. The hemispherical harmonics (HSHA) approach was first proposed by Huang et al. (2006) \cite{Huang_2006} as an extension to the original work of Brechb{\"u}hler et al. (1995) \cite{Brechbuhler_1995}. More recently, we proposed the spherical cap harmonics (SCHA) \cite{shaqfa2021b} to generalize the analysis domains from hemispheres to spherical caps. The SCHA approach analyzed open surfaces with a customizable parameterization domain (i.e., spherical caps prescribed by an opening angle $\theta_c$). The SCHA approach suffered from the slow evaluation of the associated Legendre functions, unlike the standard Legendre polynomials used in (hemi-) spherical harmonics. To efficiently study nominally flat self-affine rough surfaces, we also proposed the disk harmonics analysis (DHA) in Shaqfa et al. (2025) \citep{Shaqfa2025_DHA}. However, our most general work for open surfaces was using the hemispheroidal harmonics (HSOH) by Choi and Shaqfa (2025) \cite{Choi2025}. In the latter, we propose multiple parameterization approaches for oblate and prolate hemispheroids of arbitrary sizes. In HSOH, we use the standard associated Legendre polynomials with a linear scale onto a hemispheroid. These approaches are crucial for analyzing the morphology and roughness of surface patches, with direct implications for tribological studies.
\par
The spectral decomposition approaches consist of three main stages: (i) mapping surfaces into a reference analysis domain $\mathcal{M}$ (e.g, sphere, disk, spheroid, hemispheroid), (ii) orthogonal projection of the surface vertices onto the harmonics of the associated analysis domain $\mathcal{H}$ (i.e., mapping from physical to Fourier space), and (iii) the inverse transformation or the reconstruction stage $\mathcal{H}^{-1}$. The first two stages are excessively studied in the abovementioned literature. However, the reconstruction stage is rarely discussed, and it will be the focus of this work. Mathematically, the reconstruction stage maps the surface from Fourier space to the physical space $\mathbb{R}^3$ (i.e., inverse transformation).
\par
\begin{figure}[ht!]
    \centering
    \includegraphics[width=1.0\linewidth]{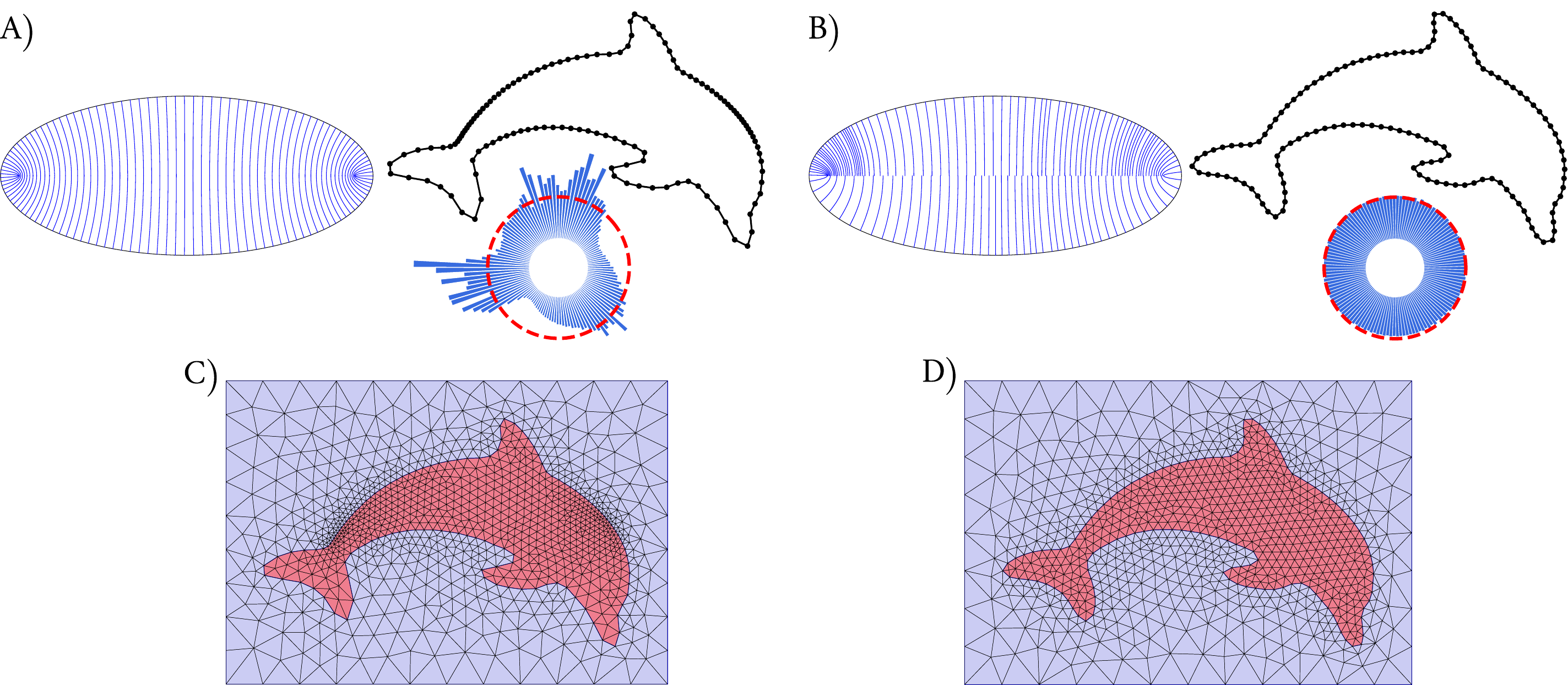}
    \caption{Reconstructing and meshing 2D contours via the elliptic harmonics approach proposed in \cite{Shaqfa2024_SOH}. (A) Uniform sampling of the hyperbolic curvilinear coordinates $\eta$ (left) and the corresponding reconstruction (right). Below is the radial distribution of the spacing between two consecutive points on the reconstructed contour, where the red dashed line represents the mean of the spacings. (B) The herein proposed diffusion-based sampling of the $\eta$ coordinates (left), where the resulting reconstruction (right) corresponds to uniformly sampled points on the contour as it converges to the mean spacings of (A). (C) The corresponding finite element (FEM) mesh of the reconstruction bulk in case (A). (D) The FEM mesh of the reconstructed contour in case (B).}
    \label{fig:sch_dolphin}
\end{figure}
\begin{figure}[ht!]
    \centering
    \includegraphics[width=0.99\linewidth]{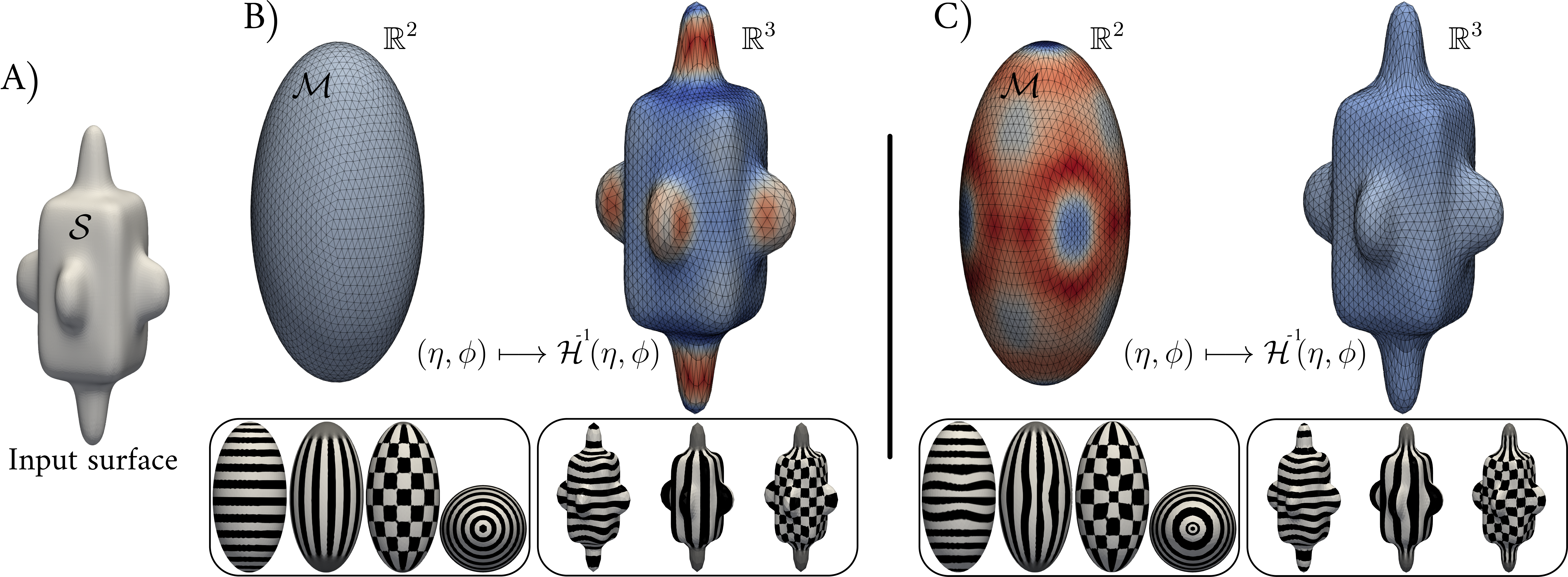}
    \caption{The difference between the reconstruction $\mathcal{H}^{-1}: \mathbb{R}^2 \longrightarrow \mathbb{R}^3$ of surfaces via uniform and nonuniform sampling of the spheroidal coordinates $(\eta, \phi)$. (A) The input surface $\mathcal{S} \subset \mathbb{R}^3$. (B) SOH reconstruction using $\eta\phi$--coordinates of a scaled icosahedron, resulting in nonuniform surface reconstruction. (C) Same as (B), but with a nonuniform sampling of $\eta\phi$--coordinates, obtained via isotropic diffusion, that results in uniform surface reconstruction. Color maps represent normalized local areas of the triangulated surfaces; blue is for small and red is for large areas relative to the mean triangular area. To visualize the duality of sampling, we introduced the gray-scale insets (below) that show the texture deformation on the spheroidal coordinates and the corresponding reconstruction.}
    \label{fig:sch_bench}
\end{figure}

The standard reconstruction approach uniformly samples the coordinates of the analysis domains \cite{Brechbuhler_1995}. For instance, in spherical harmonics (SH), spheres are represented by geodesic polyhedra (icospheres) that uniformly generate a triangulated surface of a unit sphere. However, due to the first fundamental form of the reconstruction step, the resulting surface will not be uniformly sampled. Instead, we argue that points sampling should account for the local stretch/compression on the reconstructed surface starting from the reference domains. To better visualize this effect, we first used the 2D elliptic counterpart of SOH, which was also proposed in \cite{Shaqfa2024_SOH}. Figure~\ref{fig:sch_dolphin}A shows how the reconstruction of a 2D contour from an equidistantly sampled $\eta$--coordinate, the hyperbolic sections, does not result in a uniform reconstruction of the contour, unlike the diffusion-based sampling in Fig.~\ref{fig:sch_dolphin}B. More generally, the equidistant sampling of spheroidal coordinates $(\eta, \phi)$, as shown in Fig.~\ref{fig:sch_bench}A, does not result in uniform meshes on the reconstructed surface, unlike the diffusion-sampled coordinates as shown in Fig.~\ref{fig:sch_bench}B. The herein proposed approaches can be seen as a reparameterization of the surface by resampling the local curvilinear coordinates on the analysis manifold.
\par 
Historically, the problem of uniformly sampling 2-manifolds is critical in multiple scientific domains, such as statistical sampling, numerical integration of quadrature points \cite{Gross2017}, signal processing \cite{McEwen2011}, and computer graphics \cite{Peters2023}. Multiple approaches were used to force a uniform distribution of points onto 2-manifolds, such as in \cite{pauly_2002, Damelin2003, ztireli2010, Brauchart2015}. A similar concept was recently employed in computer graphics to preserve landmarks and texture between surfaces in $\mathbb{R}^3$ and a flat map $\mathbb{R}^2$ where standard measurements are conducted. The latter are known as area-preserving maps. Choi et al. (2018) proposed a diffusion-based analogy to construct area-preserving maps between a simply connected open surface and genus-0 disks \cite{Choi2018}. Similar works on different manifolds were proposed for multiply connected disks \cite{Lyu2024_multiply}, 2-spheres \cite{Lyu2024_spheres}, ellipsoids \cite{Lyu2024_ellipsoid}, and toroids \cite{Yao2024}. The latter approaches were the main inspiration behind the current work for uniformly remeshing surfaces.
\par
This paper starts by reviewing the applications and the recent harmonic decomposition approaches in Section \ref{sec:intro}. Section \ref{sec:methods} summarizes the decomposition approaches used and the proposed isotropic and anisotropic diffusion-based remeshing. The remeshing results of closed and open surfaces and their performance were discussed in Section \ref{sec:res}. Finally, conclusions and future works are presented in Section \ref{sec:conc}.

\section{Methodology}
\label{sec:methods}
\par
In heat kernels, heat \textit{fluxes} from \textit{high} to \textit{low} temperatures to reach a steady-state condition throughout the body. Analogously, we here use the diffusion equation to redistribute mesh elementary areas, by diffusing large elements to subsequently increase the areas of smaller ones, thereby reaching a steady-state condition where the mesh is uniformly distributed (equalized) throughout the surface or contour.
\par
The herein proposed approach is suitable for all harmonic decomposition methods; however, this section focuses on both open and closed surfaces with genus-0 topologies. The domain of solutions for the diffusion problems is defined in the background coordinates of the harmonic basis; hence, spheroids and hemispheroids. Also, by \textit{mesh} we refer to a triangulated surface of 2-manifolds or linear segments of contours in 1-manifolds. With proper discrete differential operators, this section can be directly extended to any simplicial complex, e.g., quadrilateral or polygonal elements.
\par
This section provides a brief overview of the spectral decomposition of surfaces, the proposed nonlinear diffusion-based approach, the discrete differential operators, and algorithmic proposals for solving the problem of uniformly meshing (sampling) 2-manifolds.

\subsection{Spheroidal coordinates and harmonics decomposition}
\label{subsec:SOH}
\par
As this work defines the diffusion problems in spheroidal coordinates, we summarize here the orthogonal coordinates of prolate and oblate domains, the harmonic basis functions, and review the overall morphological decomposition pipeline.

\subsubsection{Spheroidal coordinates}
\par
Spheroidal coordinates are generated by revolving the orthogonal confocal elliptic coordinates $(\zeta, \phi)$ about their vertical or horizontal axes to compose oblate or prolate coordinates, respectively. The resulting orthogonal confocal coordinates $(\zeta, \eta, \phi)$ can be written in the parametric form:
\begin{equation}\label{eqn:oblate_prolate_para}
    \begin{aligned}
        \underbrace{
        \begin{aligned}
            x(\zeta, \eta, \phi) & = e \ \cosh \zeta \ \cos \eta \ \cos \phi,
            \\ y(\zeta, \eta, \phi) & = e \ \cosh \zeta \ \cos \eta \ \sin \phi,
            \\ z(\zeta, \eta) & = e \ \sinh \zeta \ \sin \eta,
        \end{aligned}
        }_{\text{Oblate spheroids}}
        \quad\quad
        \underbrace{
        \begin{aligned}
             x(\zeta, \eta, \phi) & = e \ \sinh \zeta \ \sin \eta \ \cos \phi,
            \\ y(\zeta, \eta, \phi) & = e \ \sinh \zeta \ \sin \eta \ \sin \phi,
            \\ z(\zeta, \eta) & = e \ \cosh \zeta \ \cos \eta.
        \end{aligned}
        }_{\text{Prolate spheroids}}
    \end{aligned}
\end{equation}
Similar to the spherical coordinates, $\phi \in [0, 2\pi]$ is the azimuthal angle generated from revolving the elliptic coordinates about the vertical or horizontal axes. The $\zeta \in [0, \infty[$ is a radial-like coordinate that corresponds to confocal ellipsoids in the space. Orthogonal to $\zeta$ and $\phi$, we have the latitude-like coordinate $\eta$. For oblate coordinates $\eta \in [-\pi/2, \pi/2]$ and for the prolate ones $\eta \in [0, \pi]$. In the case of hemispheroidal coordinates, considering the Northern hemispheroid, $\eta \in [0, \pi/2]$ for both oblate and prolate hemispheroids (Choi and Shaqfa (2025) \cite{Choi2025}). For completeness, in Shaqfa et al. (2024) \cite{Shaqfa2024_SOH} we show that the elliptic coordinates $(\zeta, \eta)$ can be analytically computed as:
\begin{equation}
\label{eqn:inv_coords}
\begin{aligned}
    \underbrace{
    \begin{aligned}
        \eta_{ob} &= \Im\left\{\cosh^{-1}\left( \frac{\rho_{ob} + i z}{e}\right)\right\}, \\
        \zeta_{ob} &= \Re\left\{\cosh^{-1}\left( \frac{\rho_{ob} + i z}{e}\right)\right\},
    \end{aligned}
    }_{\text{Oblate spheroids}} 
    \quad
    \underbrace{
    \begin{aligned}
        \eta_{pr} &= \Im\left\{\cosh^{-1}\left( \frac{i \rho_{pr} + z}{e}\right)\right\}, \\
        \zeta_{pr} &= \Re\left\{\cosh^{-1}\left( \frac{i \rho_{pr} + z}{e}\right)\right\}.
    \end{aligned}
    }_{\text{Prolate spheroids}}
\end{aligned}
\end{equation}
Where $\rho = \sqrt{x^2 + y^2}$. We here denote the parametric coordinate transformation in Eq.~\eqref{eqn:oblate_prolate_para} by $\mathcal{T}: (\zeta, \eta, \phi) \mapsto (x, y, z)$, and the inverse transformation $\mathcal{T}^{-1}: (x, y, z) \mapsto (\zeta, \eta, \phi)$. Additionally, we here assume that $\mathcal{T} \circ \mathcal{T}^{-1}(x, y, z)= \mathbf{id}_{(x, y, z)}$ and $\mathcal{T}^{-1} \circ \mathcal{T}(\zeta, \eta, \phi)= \mathbf{id}_{(\zeta, \eta, \phi)}$ are the identity maps. We will build on these identities to propose the pullback operator introduced in Section \ref{subsec:iso_diffuse}.

\subsubsection{Spheroidal harmonics}
\par
The confocal spheroidal coordinates have separable analytic solutions for Laplace's equation. In our applications for studying the morphology of 2-manifolds, we utilize shell-like surface harmonic bases \cite{Shaqfa2024_SOH}. As our approach depends on remeshing surfaces and contours, followed by the bulk mesh, we can assume that the radial-like $\zeta$-coordinates of the spheroidal coordinates are constant ($\zeta = \zeta_0$). Thus, the harmonic expansion of a function $f(\eta, \phi)$ can be approximated in the general form:
\begin{equation}\label{eqn:general_sol}
    f(\eta, \phi) \approx \sum_{n=0}^{N_{max}} \sum_{m=-n}^{n} N_m^n \ A_m^n \ \underbrace{P_m^n(\xi)}_{\text{ALP}} \ \overbrace{e^{i m \phi}}^{\text{Fourier}}. 
\end{equation}
From the separable solution, we obtain bases that are constituted of two parts, namely, the Associated Legendre Polynomials (ALP) of the first kind $P^n_m(\xi)$ for the $\eta$--coordinate and the Fourier basis $e^{im\phi}$ for the $\phi$--coordinate. Where $n$ and $m$ are the degrees and orders, respectively, of a given harmonic function, and $N_{max}$ is the maximum expansion degree. $N^n_m$ is a normalization factor to constitute orthonormal basis functions, and $A^n_m$ is the expansion Fourier coefficient for a given harmonic defined by $n$ and $m$. 
\par
For oblate spheroids $\xi =  \sin \eta$, and $\xi =  \cos \eta$ for prolates and spheres. In the case of hemispheroids, they are a special case with a Neumann boundary condition on the free edge. For this, we leverage affine transformations, such that $\xi = 2\sin \eta -1$ for oblate hemispheroids, and $\xi = 1 - \cos \eta$ for prolate hemispheroids or hemispheres; refer to Choi and Shaqfa (2025) \cite{Choi2025} for more details.

\subsubsection{Spectral morphological analysis}
\par
Let an arbitrary genus-0 surface, denoted by $\mathcal{S}$, be a simply connected 2-manifold $\subset\mathbb{R}^3$ with $n_v$ vertices and $n_f$ triangulated faces. Spectral methods can be used to study morphology by computing invariant shape descriptors, which are then used to compare, classify, and generate surfaces. These descriptors are computed as the power spectral density (PSD) of the expanded Fourier weights $A_m^n$ \cite{Shaqfa2024_SOH}.
\par
These spectral decompositions are conducted on a reference analysis domain $\mathcal{M}$. To analyze the morphology, we need a proper parameterization (mapping) approach to map an input surface $\mathcal{S}$ onto the corresponding analysis domain $\mathcal{M}$, such that $(x,y,z) \in \mathcal{S} \mapsto \left(\eta(x,y,z),\, \phi(x,y,z)\right) \in \mathcal{M}$. In SOH \cite{Shaqfa2024_SOH} and HSOH \cite{Choi2025}, the size of the analysis domain is prescribed with $\zeta_0$ and $e$ as degrees of freedom (DOF). These DOFs can be calculated by traditional fitting methods to minimize the topological distortions on $\mathcal{M}$.
\par
The process of finding Fourier weights is referred to as the forward spectral analysis or harmonic decomposition and is denoted by $\mathcal{H}$; refer to \cite{Shaqfa2023OnMethod} for numerical details on the decomposition methods. This process produces $\beta = (N_{max} +1)^2$ complex-valued Fourier weights $\mathbf{Q} \in \mathbb{C}^{\beta\times3}$, such that $A_{m, d}^n \in \mathbf{Q}, \forall \, d \in \{x, y, z\}$. Based on $d$, these weights can be arranged in three columns that correspond to the expansion of the parametric coordinates $x(\eta, \phi)$, $y(\eta, \phi)$, $z(\eta, \phi)$. The reconstruction process, inverse transformation, is denoted by $\mathcal{H}^{-1}$, and it is explained in Eq.~\eqref{eqn:general_sol}. However, instead of using $\beta = (N_{max} +1)^2$ basis for reconstruction, we can benefit from a faster reconstruction via the conjugate symmetry property and exclude the redundant negative orders $m < 0$. This reduces the used basis to $\hat{\beta} = (N_{max} + 1) (N_{max}+2)/2$, and the map $\mathcal{H}^{-1}$ can be instead written in the form (based on a private communication with Dr. Justin Willmert) \cite{Shaqfa2023OnMethod}:
\begin{equation}
\label{eqn:rec_conj}
\begin{array}{rcl}
    \mathcal{H}^{-1}: \mathbb{R}^2 & \xlongrightarrow{\mathbb{C}^{\beta \times 3}} &\mathbb{R}^3\\
    f(\eta, \phi) & \approx & \displaystyle\sum_{n=0}^{N_{max}} \sum_{m=0}^{n} \Re \ \Big\{ (2-\delta_{m_0}) \ N_m^n \ A_m^n \ P_m^n(\xi) \ e^{i m \phi} \Big\},
\end{array}
\end{equation}
where $\delta_{m_0}$ is Kronecker delta, such that $\delta_{m_0} = 1$ when $m = 0$, otherwise $\delta_{m_0} = 0$. As it is evaluated once per time step, using Eq.~\eqref{eqn:rec_conj} instead of \eqref{eqn:general_sol} for reconstruction significantly reduces the runtime of the diffusion-based approach.

\subsection{Isotropic remeshing by diffusing the curvilinear \texorpdfstring{$\eta\phi$}{}--coordinates}
\label{subsec:iso_diffuse}
\par
Once reconstructed with harmonic decomposition, we want to equalize triangular faces across the whole surface; hence, this approach is classified as area-preserving \cite{Jin2004}. This approach differs from traditional remeshing ones in that it preserves the morphological features of the surface, i.e., maintaining constant Fourier weights. In other words, this diffusion approach can be seen as a reparameterization or resampling of the curvilinear coordinates on the analysis domain to ensure uniform reconstruction. 
\par
To diffuse the vertices (coordinates) of $\mathcal{M}$, we define a per-vertex scalar field that represents the area density derived from the reconstructed mesh in $\mathbb{R}^3$. The scalar density function is denoted by $u(\eta, \phi) = A_i / \sum_\Omega A_i$ and represents the normalized Voronoi area on the $i^\text{th}$ vertex. When the scalar field $u_t(\eta_t, \phi_t)$ is diffused on $\mathcal{M} \subset\mathbb{R}^2$ 2-manifolds, the $(\eta, \phi)$ coordinates will undergo continuous deformation on the analysis domain $\mathcal{M}$ to minimize the variation in mesh area density in $\mathbb{R}^3$ once reconstructed.
\par
The diffusion operators depend on the intermediate diffeomorphisms coordinates of $\mathcal{M}_t$ at each time step $t$, making this problem inherently nonlinear. A steady state solution of this nonlinear problem converges to a diffeomorphism $\mathcal{M}_\infty$ that corresponds to a uniformly sampled mesh on the reconstructed surface, i.e., the density function $u_{\infty}(\eta_\infty, \phi_\infty) = constant$. Solving this problem without changing the size of the analysis domain and Fourier weights preserves the overall area and volume of the reconstructed surface (i.e., a satisfies a physical constraint).
\par
Mathematically, the diffusion problem can be expressed as a parabolic partial differential equation (PDE), and it can be written in the form:
\begin{equation}
    \label{eq:diffusion}
    \begin{aligned}
        \partial_t u &= -\nabla \cdot \left( -\mathbf{D} \, \nabla u \right) \quad &&\text{on} \quad \Omega, \\
        \left( -\mathbf{D} \, \nabla u \right) \cdot \mathbf{n_e} &= g(u) \quad &&\text{on} \quad \partial\Omega.
    \end{aligned}
\end{equation}
Where $\partial_t$ is a derivative with respect to time and $\mathbf{D}$ is the diffusion tensor. For a closed 2--manifold of spheroids, the coordinates are periodic, which means that $\partial \Omega = \emptyset$. For single-edged open surfaces, see Section \ref{subsec:res_iso_open}, $\mathbf{n_e}$ is the normal vector of the edge, and $g$ is the Neumann boundary flux. In this section, we consider an isotropic diffusion where the diffusion tensor is constant across all directions and parallel to the gradient of the area density field; that is, $\mathbf{D} \nabla u$ is parallel to $\nabla u$. So, we here consider $\mathbf{D} = \mathbf{I}$ (the identity matrix), and the isotropic Laplace-Beltrami operator can be simplified to:
\begin{equation}
    \nabla^2 u = \nabla \cdot (\nabla u).
\end{equation}
The discrete form of this Laplace-Beltrami operator can be realized from the standard weak form of the diffusion problem. To account for the nonlinearity of the diffusion problem and the evolution of the $(\eta, \phi)$ coordinates, the Laplace-Beltrami operator $\mathbf{L}_{iso}\left(\mathcal{T}(\eta_t, \phi_t) \right)$ must be recomputed each time step $t$. Hence, we use the efficient cotangent formula \cite{Meyer2003, libigl} for the operator on $\mathcal{M}$, and it can be written as:
\begin{equation}\label{eqn:laplacian_iso}
    \mathbf{L}_{iso}\left(\mathcal{T}(\eta, \phi) \right) = \int_{\Omega} -\nabla u^T \, \nabla u \, d\Omega = -\mathbf{G}^T\,\mathbf{A}\,\mathbf{G},
\end{equation}
where $\mathbf{G} \in \mathbb{R}^{3 n_f \times n_v}$ is the Cartesian gradient operator, and $\mathbf{A} \in \mathbb{R}^{3 n_f \times 3 n_f}$ is the diagonal lumped mass matrix. The resulting isotropic Laplace-Beltrami operator $\mathbf{L}_{iso} \in \mathbb{R}^{n_v \times n_v}$ is a sparse symmetric matrix.
\par
Due to the mismatch between the tangent spheroidal domain $\mathcal{M} \subset \mathbb{R}^2$ of the diffusion problem and the Cartesian embedding of the Laplacian operator in Eq.~\eqref{eqn:laplacian_iso}, the diffused coordinates deviate from the tangent surface $\mathcal{M}$. This post-diffusion perturbed domain $\mathcal{M}_{\varepsilon}$ can be seen as a noisy image of the original $\mathcal{M}$, where the $\zeta$--coordinate can be expressed as:
\begin{equation*}
    \zeta_\varepsilon = \zeta_0 + \varepsilon.
\end{equation*}
To correct this behavior after each diffusion step, we introduce a \textit{pullback} operator that projects the points of $\mathcal{M}_\varepsilon$ back to the tangent surface $\mathcal{M}$. For this, we use the nonlinear hyperbolic projection from \cite{Shaqfa2024_SOH} as a pullback operator $\phi^*: \mathcal{M}_\varepsilon \longrightarrow \mathcal{M}$.
When $\varepsilon \in \mathbb{R}$ is a small post-diffusion perturbation, we can assume that:
\begin{equation*}
    \mathcal{T (\zeta_\varepsilon, \, \eta + \delta_\eta, \, \phi + \delta_\phi)} \approx \mathcal{T}(\zeta_0, \, \eta + \delta_\eta, \, \phi + \delta_\phi),
\end{equation*}
such that $\delta_\eta$ and $\delta_\phi$ are post-diffusion steps for the $(\eta, \phi)$--coordinates on $\mathcal{M}$ while $\zeta_0$ is constant. For large diffusion time steps, $\phi^*$ can cause the triangulated mesh to flip normals (i.e., result in self-intersected meshes). Hence, we introduced an adaptive time stepping that reduces the current time step upon detecting any flip of normals.
\par
The weak form of the diffusion problem can be reduced to a linear algebraic system where the time derivative is approximated at the $i^{\text{th}}$ time step as $\partial_t u \approx (u_{i+1} - u_i) / \Delta t$. Here, $\Delta t$ is the largest stable time step. Using the standard backward Euler scheme, implicit integration, we can solve the algebraic system:
\begin{align}
    \mathbf{A}_i\,\frac{u_{i+1} - u_i}{\Delta t} &= \mathbf{L}_{i}\, \ u_{i+1}, \\
    (\mathbf{A}_i - \Delta t\, \mathbf{L}_{i})\, u_{i+1} &= \mathbf{A}_i \,u_{i}.
    \label{eqn:discrete_form}
\end{align}
\par
Algorithm \ref{algo:iso} summarizes the pseudo-code of the nonlinear diffusion solver implemented in this work. As it is important for the stability of the numerics, we scale all the input surfaces such that the total surface area is unity. This scale can be reversed by the end of the remeshing approach. For efficient computations, we used a sparse representation of all the matrices and operators. The resulting sparse system was solved using the Generalized Minimal Residual Method (GMRES) iterative solver with an Algebraic Multigrid (AMG) preconditioner based on the Ruge–St\"uben method; see details in the PyAMG library \cite{pyamg2023}. The most computationally demanding parts of the critical loop of Algorithm \ref{algo:iso} are: (i) the Laplacian assembly and (ii) the surface reconstruction step in Eq.~\eqref{eqn:rec_conj}. To address the numerical complexity of the reconstruction problem, Section \ref{sec:res} discusses a stage-based strategy that drops the total time by at least $50\%$, assuming a single-core vectorized implementation of the algorithm.
\par
In principle, implicit solving schemes are unconditionally stable. However, as the diffusion problem here is nonlinear and continuously deforms the mesh, using large time steps causes the face orientation to flip due to the large jumps in the pullback operator. Moreover, $\Delta t$ directly affects the condition number of the sparse matrix $\mathbf{A}_i - \Delta t \ \mathbf{L}_{iso}$; thus, affecting the overall convergence time of the GMRES solver.

\subsubsection{Imposing Neumann boundary condition for open surfaces}
\label{subsubsec:openBC}
\par
In this section, the goal is to achieve a stable nonlinear diffusion process for remeshing open surfaces while maintaining the physical constraint of preserving the overall surface area and boundary length within acceptable error margins. For open surfaces reconstructed with the hemispheroidal harmonics (HSOH) \cite{Choi2025} and the disk harmonics (DH) \cite{Shaqfa2025_DHA} approaches, using Neumann boundary condition (BC) $\partial_t \ u = 0, \ \forall \ u \in \partial \Omega$ requires careful numerical treatment. Choi et al. \cite{Choi2018} proposed a boundary projection algorithm to correct the numerical flow of the boundary vertices of a flat disk with a no-flux BC. With the proposed pullback operator $^*\phi$ (line $14$ in Algorithm \ref{algo:iso}), the projection behavior is de facto covered. In the HSOH approach, however, singular values can be obtained by evaluating the basis functions on the \textit{true} boundary, and numerical overflow can cause the reconstruction of the near-edge area to blow up \cite{Choi2025}. To account for this singularity in the reconstruction step, we avoided sampling points that fall exactly on the true boundary and replaced it with an edge that falls within the neighboring boundary layer; hence, shifting back the $\eta$--coordinate on $\Omega$ with a small number $\varepsilon_\eta$; see \cite{Choi2025}.
\par
Applying Neumann BC for the herein mesh diffusion problem without the true edge explicitly presented needs additional treatment. To preserve the physical fidelity of the diffusion system and compensate for shifting the true edge, equivalent \textit{artificial} boundary conditions (ABC) can be applied on the new edge. This shifted edge can be written as $\partial \Omega_\varepsilon = \left\{ (\eta, \phi) \in \mathbb{R}^3 \,\middle|\, \eta = \frac{\pi}{2} - \varepsilon_\eta,\ 0 < \phi \leq 2\pi \right\}$. When we remesh closed surfaces, $\partial \Omega = \emptyset$, the diffused mesh converges to the mean triangular face area of the mesh at $t=0$. So, we imposed a point-wise flux-averaged Neumann ABC that is equivalent to the current mean area of the solution, such that $\partial_t\ u_{i} = \bar{u}_{i-1},\ \forall \ u \in \partial \Omega_\varepsilon$. This logic is similar to the flux-averaged approach presented in \cite{Parker1984}. The results in Section \ref{sec:res} suggest that this approach is reliable for obtaining stable solutions while preserving physical constraints.

\begin{algorithm}[H]
\label{algo:iso}

\caption{Diffusion of the $\eta\phi$--coordinates for equalizing surface meshes}

\KwIn{Initial triangulation (sampling) of $\mathcal{M}_0 (\eta, \phi)$, Fourier weights $\mathbf{Q} \in \mathbb{C}^{(N_{max}+1)^2 \times 3}$, iterations $I_{max}$}
\KwOut{Adaptive sampling of $\mathcal{M}_{\infty}$ and a uniform mesh of $\mathcal{H}^{-1}_{\infty}(\eta_{\infty}, \phi_{\infty})$}

Initialize memory\;
Normalize the input surface area to be $1.0$\;
Set $u \leftarrow u_0$\;
Set $t \leftarrow 0$\;
Compute $\Delta t$\;

\While{$i \leq I_{max}$}{
    Initialize nonlinear iteration: $u_{(0)} \gets u$\;

    Set $(\eta_i, \phi_i) \leftarrow (\eta_{i-1}, \phi_{i-1})$\;

    Compute the Laplace-Beltrami $\mathbf{L}_{i} (\eta_i, \phi_i)$ and Mass $\mathbf{A}_i(\eta_i, \phi_i)$\;

    Apply Neumann boundary conditions if $\partial \Omega \neq \emptyset$\;

    Compute $u_{i}$ from implicitly solving Eq.~\eqref{eqn:discrete_form}\;

    Compute the Cartesian gradient $\mathbf{G} u_i$\ averaged on vertices\;
    
    
    Update $(\eta_{i}, \phi_{i}) \leftarrow (\eta_i, \phi_i) + \mathcal{T}^{-1}(\mathbf{G} u_i)$\;

    Apply the pullback operator $^*\phi: \mathcal{M}_{\varepsilon, i} \longrightarrow \mathcal{M}_i$\;
    
    Check for sign flips of normals on $\mathcal{M}_i(\eta_i, \phi_i)$\;

   \If{Normals flip}{
        \tcp{Reduce $\Delta t$ and repeat same iteration}
        $\Delta t \gets \Delta t / 2$\;
        \textbf{continue}\;
    }
    Reconstruct surface $\mathcal{H}^{-1}(\eta_i, \phi_i)$\;

    Update the area errors of $u_{i+1}(\eta_i, \phi_i)$ from $\mathcal{H}^{-1}(\eta_i, \phi_i)$\;

    Update the time step $\Delta t$ $ \leftarrow \mathcal{M}_{i}$\;
    
    $i \coloneq i + 1$\;
}
Rescale the output $\mathcal{H}^{-1}_{\infty}$ to retain its original surface area\;
\end{algorithm}

\subsubsection{Anisotropic Laplacian operator}
\label{subsec:aniso_diffuse}
\par
The herein isotropic diffusion approach can be classified as an area-preserving algorithm that does not preserve the interior angle structure of the mesh (i.e., it is non-conformal) \cite{Jin2004}. These approaches may result in distorted triangles with obtuse angles, especially near large surface protrusions. For numerical simulations, obtuse-angle triangles are not desired as they can be a source of numerical errors and stability problems \cite{babuvska1976}. To alleviate these distortions, we propose a modified anisotropic Laplacian operator that slows down the diffusion rate along the excessively elongated direction of the triangle and instead favors diffusivity along the perpendicular short direction.
\par
\begin{figure}[ht!]
    \centering
    \includegraphics[width=0.7\linewidth]{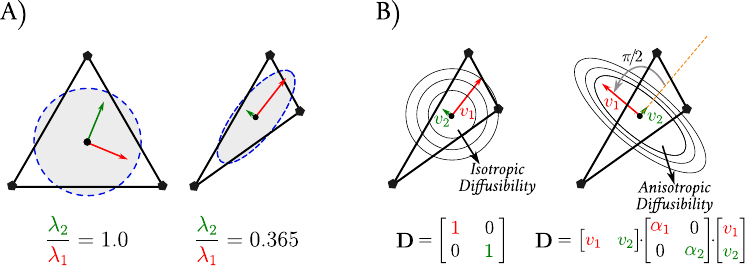}
    \caption{Schematics for the anisotropic diffusion kernel. (A) The principal distortion directors (PDD) of equilateral and non-equilateral triangles, and (B) the corresponding isotropic and anisotropic diffusibility of distorted triangles.}
    \label{fig:tri_sch}
\end{figure}

The anisotropic discrete Laplace-Beltrami operator is typically expressed as:
\begin{equation}\label{eq:aniso_laplacian}
    \mathbf{L}_{aniso} = \int_{\Omega} -\nabla u^T\, \mathbf{D}\, \nabla u \, d\Omega = - \mathbf{G}^T\,\mathbf{D}\,\mathbf{A}\,\mathbf{G},
\end{equation}
where, the positive definite diffusion tensor $\mathbf{D}$ is written as:
\begin{equation}
    \mathbf{D} = \mathbf{R}\,\Lambda\,\mathbf{R}^T.
\end{equation}
Figure \ref{fig:tri_sch}A shows two examples of triangles: the optimal (i) equilateral triangle (left) and the undesired (ii) angle-obtuse triangle (right). To measure the anisotropy of a distorted triangle, we define the aspect ratio $\text{AR} = \lambda_2/\lambda_1$ as the ratio of the smallest nonzero eigenvalue $\lambda_2$ to the largest one $\lambda_1$ of the triangle vertices $\mathbf{V}_T = [v_{t,1}, v_{t,2}, v_{t,3}]^T \ \in \mathbb{R}^3$. The rotation matrix $\mathbf{R} = [v_1^\perp, v_2^\perp]$ consists of the first $v_1$ and second $v_2$ eigenvectors of $\mathbf{V}_T$ that are rotated by $\pi/2$ counterclockwise about the normal of the triangular face. The vectors $v_1$ and $v_2$ are here called the principal distortion directors (PDD), and $\Lambda$ is a diagonal matrix of the corresponding diffusion rates, where $\Lambda = \text{diag} (\alpha_1, \alpha_2)$. The coefficients $\alpha_1$ and $\alpha_2$ are functions of the ratio $\text{AR}$, such that
\begin{align}\label{eqn:aniso_diff_param}
    \alpha_1 = e^{\left(1-\frac{\lambda_1}{\lambda_2}\right) \frac{1}{\gamma}}, \quad
    \alpha_2 = e^{\left(1-\frac{\lambda_2}{\lambda_1}\right) \gamma}.
\end{align}
The parameter $\gamma$ is introduced to control the strength of the tensor $\mathbf{D}$. For an equilateral triangle, when $\lambda_1 = \lambda_2$, we have $\alpha_1 = \alpha_2 = 1$, resulting in a unity diffusion tensor $\mathbf{D} = \mathbf{I}$, which corresponds to the isotropic case. For the case when $\lambda_1 \geq \lambda_2$, we strengthen the diffusability along the thin direction ($2^{\text{nd}}$ PDD) of the triangle, and vice versa.
\par
To efficiently compute the principal distortion directors (PDD) for each triangular face, we use the singular value decomposition (SVD) approach, such that the face vertices can be expressed as $\mathbf{V}_T = \mathbf{U}\, \Sigma\, \mathbf{V}$. This decomposes the face vertices into three directors that correspond to the eigenvalues $\Sigma = \text{diag}(\lambda_1, \lambda_2, \lambda_3)$. As $\mathbf{V}_T$ is a planar triangular face, $\lambda_3 = 0$ corresponds to the face's normal direction ($\text{N} = v_3$). The tangent PDD vectors $v_1$ and $v_2$ determine the major and minor directors of distortion, respectively; see Fig.~\ref{fig:tri_sch}. In this work, the rotation matrices are defined to rotate the surface tangents about the normals of the faces, such that $\mathbf{R}^{T} \cdot \mathbf{R} = \mathbf{I}$. Using Rodrigues' rotation matrix, $\mathbf{R} = \mathbf{I} + \bar{\mathbf{N}} \sin{\alpha} + \bar{\mathbf{N}}^2 (1 -\cos{\alpha})$, where $\alpha$ is the counterclockwise wise rotation about the face normal $\mathbf{N}$ and $\bar{\mathbf{N}}$ is written as:
\begin{equation}
\label{eq:rot_mat}
    \bar{\mathbf{N}} = \begin{bmatrix}
        0   & -N_z & N_y \\
        N_z & 0    & -N_x\\
        -N_y& N_x  & 0   \\
    \end{bmatrix}.
\end{equation}
Since $\alpha = \pi/2$, Rodrigues' rotation matrix reduces to: $\mathbf{R} = \mathbf{I} + \bar{\mathbf{N}} + \bar{\mathbf{N}}^2$.  
\par
Here, the condition number of the sparse matrix $\mathbf{A}_i - \Delta t \ \mathbf{L}_{aniso, \ i}$ is always higher than the isotropic case. Moreover, the corresponding stable time step is generally lower; one way to estimate this is by assuming $\Delta t_{aniso} = \Delta t_{iso} / \text{max}(\alpha_1)$.

\section{Results and discussions}
\label{sec:res}
In this section, we discuss the results of the proposed morphology-preserving remeshing approach. We begin by introducing a comprehensive benchmark that discusses the qualitative metrics used, the computational performance, and the proposed hierarchical scheme. Then, we consider remeshing open single-edged surfaces, followed by comparing benchmarks for isotropic and anisotropic diffusion algorithms of closed surfaces. For engineering applications, we remesh a 2D particulate microstructure of a concrete sample and a 3D one of a real stone masonry wall.

\subsection{Isotropic remeshing of closed genus-0 surfaces}
\label{subsec:res_iso_closed}
\par
\begin{figure}[!ht]
    \centering
    \includegraphics[width=1.0\linewidth]{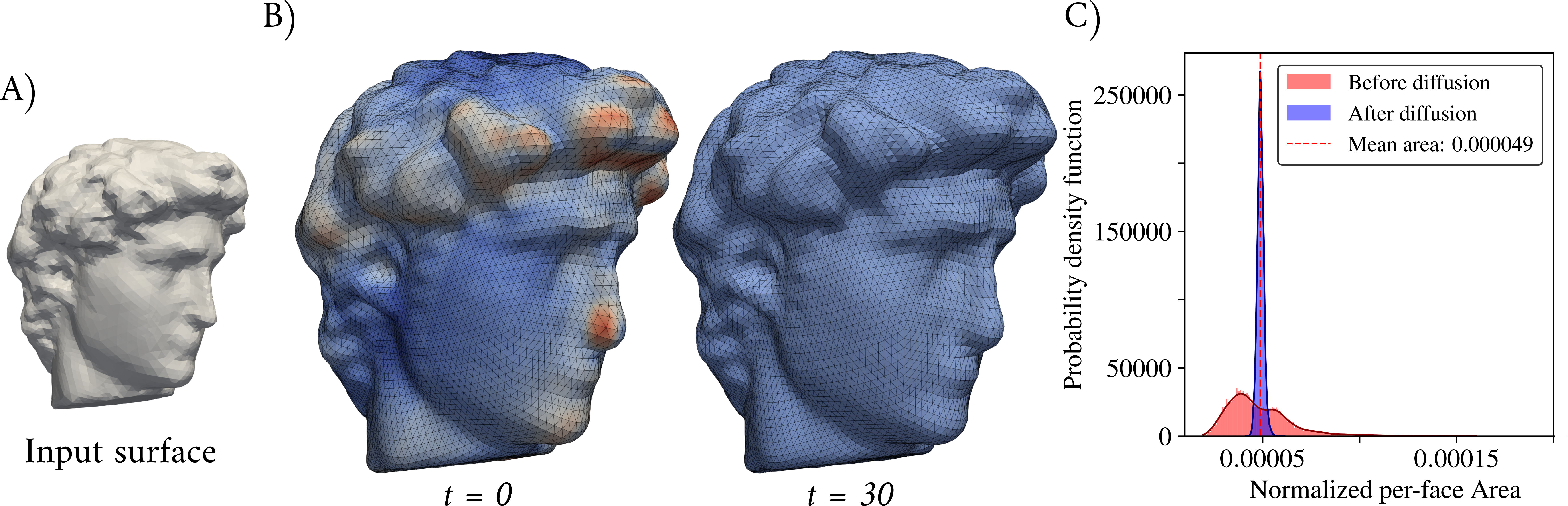}
    \caption{Isotropic diffusion for remeshing David's head bust via SOH. (B) We used five refinement cycles of an icosphere for the reconstruction process with $N_{max} = 50$ and $I_{max} = 30$. The color maps in (B) reflect the area density on the surface (red for high errors, and blue for relatively small errors). (C) Shows the PDF of the area density before (in red) and after (in blue) the diffusion.}
    \label{fig:david_diff}
\end{figure}

We first remeshed David's head bust benchmark surface, which is a closed genus-0 input surface consisting of $10,671$ vertices and $21,338$ faces (see Fig.~\ref{fig:david_diff}). With the SOH decomposition, we used $N_{max} = 50$ and reconstructed the surface using a rescaled icosahedron with five refinement cycles that correspond to $10,242$ vertices and $20,480$ faces as shown in Fig.~\ref{fig:david_diff}B ($t=0$). The color map represents the area density on the surface, with regions of high stretch density highlighted in red (e.g., protrusions like the nose tip and hair tufts), and contracted areas shown in blue (e.g., near spheroidal poles such as the top of the head). After applying the isotropic diffusion on the curvilinear coordinates, Fig.~\ref{fig:david_diff}B ($t=30$), we obtained a uniform mesh throughout the whole surface.
\par
To quantitatively measure the uniformity of the newly-sampled coordinates, Fig.~\ref{fig:david_diff}C compares the probability density functions (PDFs) of the area density before (in red) and after (in blue) the diffusion. The post-diffusion PDF shows a narrow PDF concentrated about the mean triangular area of the surface at $t=0$. Convergence to the mean value is a well-known result for diffusion problems, where no \textit{temperature} fluxes in or out of a perfectly insulated medium occur. To monitor the convergence of the system, we used the standard deviation (STD) of the area density at each time step $t$; the STD decreases over time, indicating a narrowing PDF.
\par
\begin{figure}[!ht]
    \centering
    \includegraphics[width=1.0\linewidth]{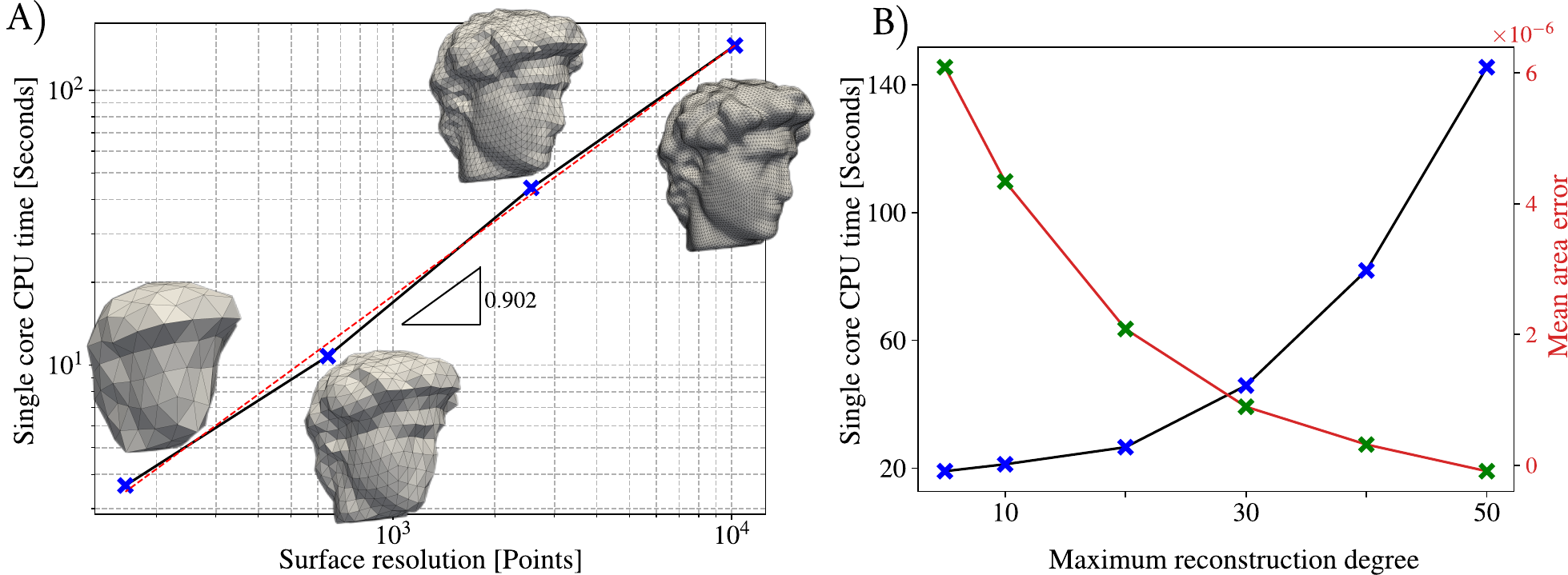}
    \caption{Computational time comparisons of different diffusion and reconstruction configurations. (A) Compares the CPU time versus the surface resolution (icosahedral refinements: $\{2, 3, 4, 5\}$) with fixed $N_{max} = 50$. (B) Compares the CPU time versus $N_{max}$ used throughout the diffusion iterations $N_{max} \in \{5, 10, 20, 30, 40, 50\}$. The same inset shows the error in the converged mean area density as a function of $N_{max}$.}
    \label{fig:david_time}
\end{figure}

Figure \ref{fig:david_time}A shows the single-core CPU time scaling with the icosahedral refinements using $I_{max} = 30$. As each surface subdivision cycle splits existing triangles into three smaller ones, the degrees of freedom of the diffusion problem quickly grow. Thus, the solving time increases; however, this is still not the bottleneck of this approach. The surface reconstruction $\mathcal{H}^{-1}_t$ evaluated at each time step, a function of $N_{max}$ in Eq.~\eqref{eqn:rec_conj}, grows by $(N_{max}+1)(N_{max}+2)/2$ \cite{Shaqfa2023OnMethod}. This makes the reconstruction complexity the main bottleneck in this work. To visualize the complexity of Eq.~\eqref{eqn:rec_conj}, Fig.~\ref{fig:david_time}B shows the time versus $N_{max}$ of the diffusion problem in Algorithm \ref{algo:iso}. The CPU time quadratically increases from about $20$ seconds for $N_{max} = 10$ to roughly $140$ seconds for $N_{max} = 50$, over the same diffusion setting $I_{max} = 30$. On the same figure, with increasing $N_{max}$, the mean area density error (absolute percentage) decreases quadratically relative to the mean area density reached when $N_{max} = 50$. This is expected since not all reconstruction degrees are included in equalizing the mesh when $N_{max} < 50$.
\par
\begin{figure}[!ht]
    \centering
    \includegraphics[width=1.0\linewidth]{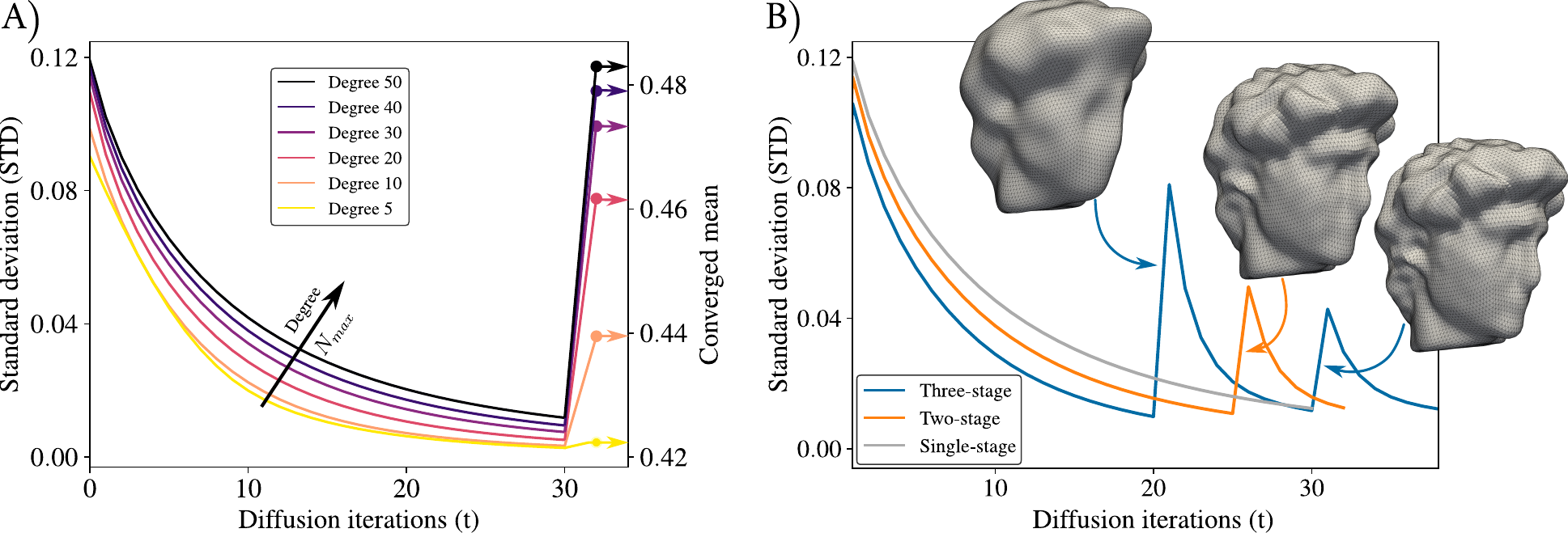}
    \caption{The effect of $N_{max}$ on the convergence of the isotropic diffusion problem. (A) The convergence rates associated with $N_{max}$ using icosahedral with five refinement cycles. For different $N_{max}$, we converge to slightly different mean densities than the mean of the input surface. (B) Multiple hierarchical solutions for the isotropic diffusion problem. A single-stage solution used $N_{max} = 50$ and $I_{max} = 30$ took $121.83$ seconds to converge to an STD of $0.01236$, the two-stage solution with $N_{max} = \{30, 50\}$ and $I_{max} = \{25, 7\}$ took $58.08$ seconds to reach an STD of $0.01256$, and three-stage solution with $N_{max} = \{15, 35, 50\}$ and $I_{max} = \{20, 10, 8\}$ took $53.96$ seconds to converge to an STD of $0.01218$.}
    \label{fig:david_stages}
\end{figure}

Morphologically, small harmonics are responsible for reconstructing large surface features, making their contributions to the first fundamental form more significant than high-frequency harmonics \cite{Shaqfa2023OnMethod}. Figure \ref{fig:david_stages}A shows the rates of convergence considering different reconstruction degrees. Lower $N_{max}$ converges faster to a lower mean density value with a significantly shorter diffusion time (with five mesh refinements and $I_{max} = 30$). This inspired us to propose a hierarchical diffusion scheme that solves the diffusion problems on multiple stages, where each stage uses a different $N_{max}^{(j)}$ and maximum number of iterations $I_{max}^{(j)}$. With this, we can reduce the majority of errors at lower $N_{max}$, where we use most of the diffusion iterations, and for higher $N_{max}$, we use fewer iterations. Figure \ref{fig:david_stages}B compares three diffusion strategies of the curvilinear coordinates to achieve somewhat similar area densities at the end. The single-stage solution takes about two minutes, while the two- and three-stage solutions take less than one minute to achieve a relatively close STD, comparable to that of the one-stage solution. The jumps on the STD curves mark the beginning of stages at which we increase $N_{max}$, as it adds to the first fundamental form. For large microstructures with multiple particles, such stages can be calibrated on one sample and then generalized for other surfaces.

\subsection{Isotropic remeshing of open genus-0 surfaces}
\label{subsec:res_iso_open}
\par
In Section \ref{subsubsec:openBC}, we explained the logic behind imposing an averaged flux on the artificial boundary condition (ABC). To test this, we recall the Matterhorn mountain benchmark, which is a genus-0 open surface from \cite{Shaqfa2025_DHA}. To monitor the stability of the obtained solutions and their physical fidelity, we monitored the change in mean area density throughout all the diffusion iterations. We also computed the relative error of the change in the outer edge length relative to the original reconstruction.
\par
\begin{figure}[!ht]
    \centering
    \includegraphics[width=1.0\linewidth]{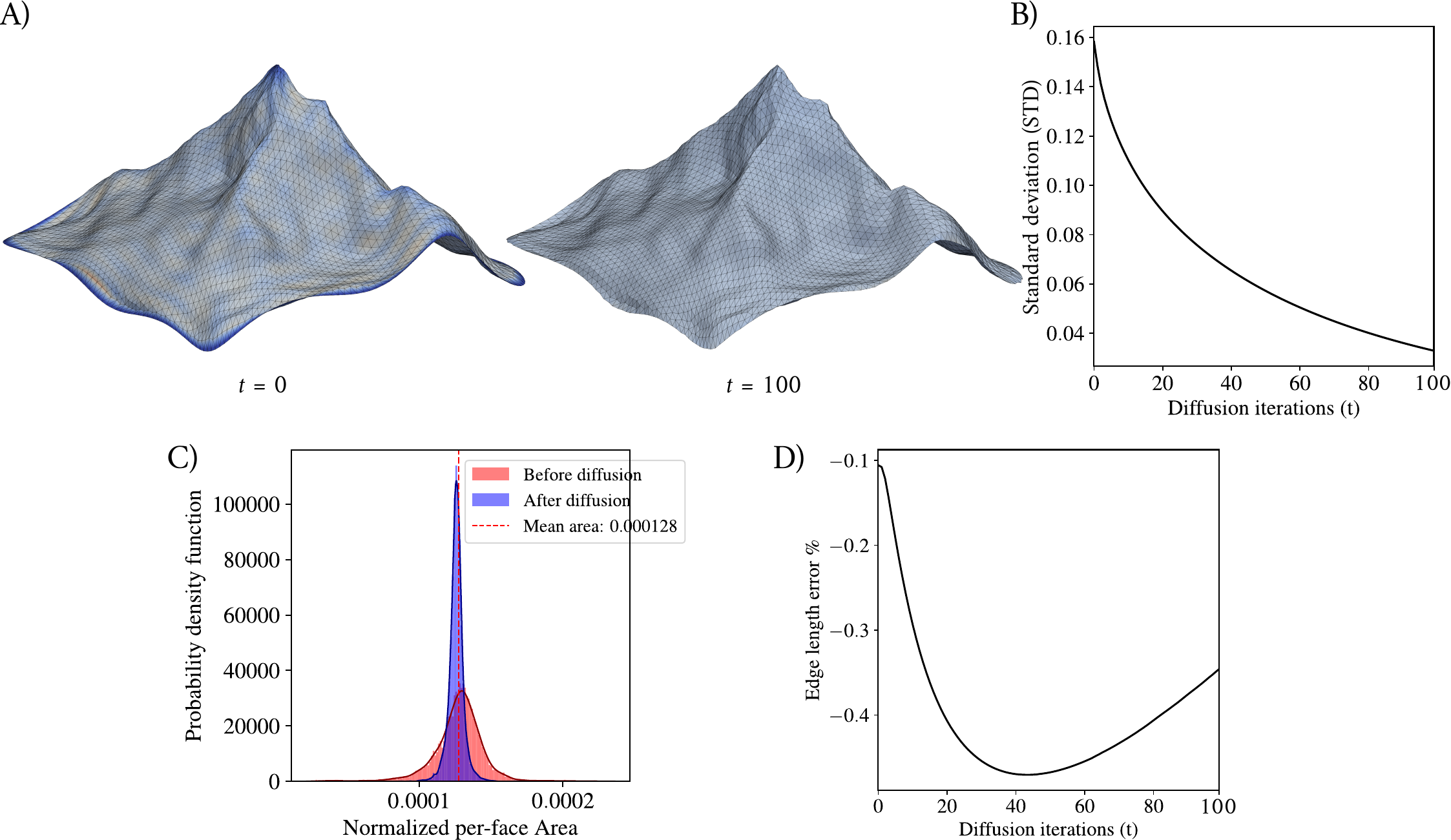}
    \caption{Remeshing of the Matterhorn open surface benchmark \cite{Shaqfa2021}, reconstructed with $4,023$ vertices and $7,839$ faces with $N_{max} = 25$. (A) The initial surface reconstruction at $t=0$ and the remeshed surface via the isotropic diffusion process at $t=100$. (B) The STD convergence versus diffusion time $t$. (C) The PDFs of the triangular faces before and after the diffusion. (D) The changes in the boundary edge length with time.}
    \label{fig:matterhorn_res}
\end{figure}

Figure \ref{fig:matterhorn_res}A, shows a comparison between the initially reconstructed surface (i.e., pre-diffusion at $t=0$) and after the coordinate diffusion at $t=100$. The results show significant improvement in the uniformity of the triangulated surface, where the STD of the area density drops exponentially with time (see Fig.~\ref{fig:matterhorn_res}B). Figure~\ref{fig:matterhorn_res}C compares the PDF of the area densities before and after the diffusion process. As a result of imposing a nonlinear averaged flux on the edge $\partial \Omega_\varepsilon$, we see that the error of the edge length is less than $-0.4\%$ (shrinkage) of the original edge length (Fig.~\ref{fig:matterhorn_res}D). Accordingly, the problem converged to a slightly different mean of surface areas of $0.000126$ instead of $0.000128$ (error of $1.6\%$ in the mean surface area). With these results, the difference between the input and remeshed surfaces is not visually noticeable, and these error margins are deemed acceptable for remeshing purposes.

\subsection{Anisotropic remeshing of benchmark surface}
\label{subsec:res_aniso_closed}
\par
The isotropic diffusion algorithm does not generally preserve the interior angular structure of the surface and can introduce obtuse triangles near large protrusions in meshes. We recall the benchmark surface in Fig.~\ref{fig:sch_bench}A, where obtuse triangles appear near the neck of these bumps. As it is widely used to quantify the shape quality of triangular faces, we propose using the traditional circumradius measure \cite{shewchuk2002}. Geometrically, the circumradius $\rho$ is the radius of a circumscribing circle for the triangular face, where large values indicate thin triangles with an obtuse angle. As equilateral triangles are considered the golden standard in FEM, we propose a normalization factor of $\sqrt{3}/a_{avg}$ to indicate how far a triangular face deviates from an equilateral one, where $a_{avg}$ is the average edge size of a face. The normalized circumradius $\hat{\rho}$ for a nondegenerate triangle lies in the range $1 \leq \hat{\rho} < 2$. For an equilateral triangle, $\hat{\rho}=1$ and $\hat{\rho} \to 2$ for a collapsed triangle.
\par
\begin{figure}[!ht]
    \centering
    \includegraphics[width=1.0\linewidth]{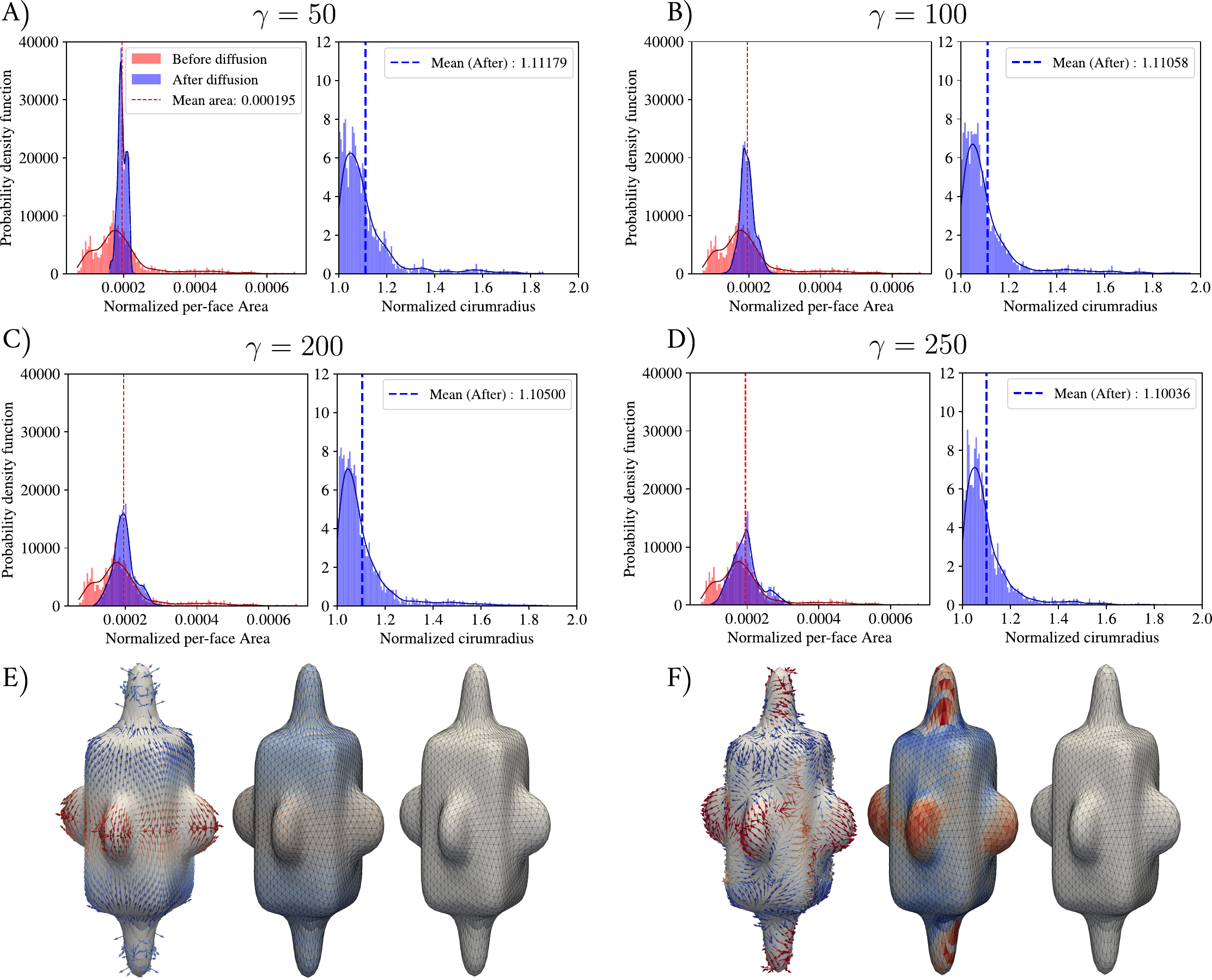}
    \caption{Comparison between different diffusion anisotropy strength $\gamma$. (A)--(D) For multiple $\gamma$ values, we compare the distortions in areas (left insets) and the circumradii (right insets) in each case. (E) Shows the flux flow (colors for magnitude) in the case of isotropic diffusion (left). The middle and left insets are for the corresponding per-face area and surface mesh of the steady-state solution with $I_{max} = 50$. (F) Same as (E), but for anisotropic diffusion with $\gamma = 250$ and $I_{max} = 100$. All color maps presented share the same scale: red for large values and blue for small ones.}
    \label{fig:anisotropic_remeshing}
\end{figure}

To alleviate the obtusity in remeshing triangular faces, we leverage the operator in Eq.~\eqref{eq:aniso_laplacian} and test different $\gamma$ values to control the degree of anisotropy in Eq.~\eqref{eqn:aniso_diff_param}. We monitored the PDFs of the resulting area densities and the normalized circumradii $\hat{\rho}$. Figure \ref{fig:anisotropic_remeshing}A--D shows the change in PDFs as a function of $\gamma$; left insets for area densities and right ones for circumradii. In general, when we favor anisotropy (large $\gamma$), we get better circumradii and less uniformity across the elementary areas. Meaning, the global measure of the mean circumradius gets closer to unity (equilateral triangles); however, the STD of the normalized area densities gets larger (less equalized). This compromise between conformalized anisotropic diffusion and area-preserving isotropic one is similar to the duality between angle-preserving and area-preserving maps in the mapping literature (see \cite{Choi2025}). Figure~\ref{fig:anisotropic_remeshing}E--F shows the comparison between isotropic and anisotropic cases, where the flux gradients show how the local directors can influence the diffusion.

\subsection{Remeshing realistic microstructures}
\label{subsec:res_masonry}
\par
With advances in 3D scanning technologies of engineering microstructures, constructing spectrally accurate geometries and discretizations from scanned data is becoming increasingly necessary to conduct detailed numerical simulations. The herein proposed remeshing methodology can be applied to any particulate inclusion microstructure, such as stone masonry walls, concrete, and metallic microstructures. We here uniformly discretize two samples: (i) a 2D concrete sample and (ii) a 3D stone masonry microstructure acquired from a digital twin (DT) scan. It is worth mentioning that we here focus on remeshing the exterior contours and surfaces of microstructures. For generating the bulk meshes, we use specialized tools, such as Gmsh \cite{Geuzaine2009} and TetGen \cite{Si2015_tetgen}, which take the exterior surfaces and contours as inputs.

\subsubsection{Remeshing 2D concrete microstructure}
\par
\begin{figure}[!ht]
    \centering
    \includegraphics[width=1.0\linewidth]{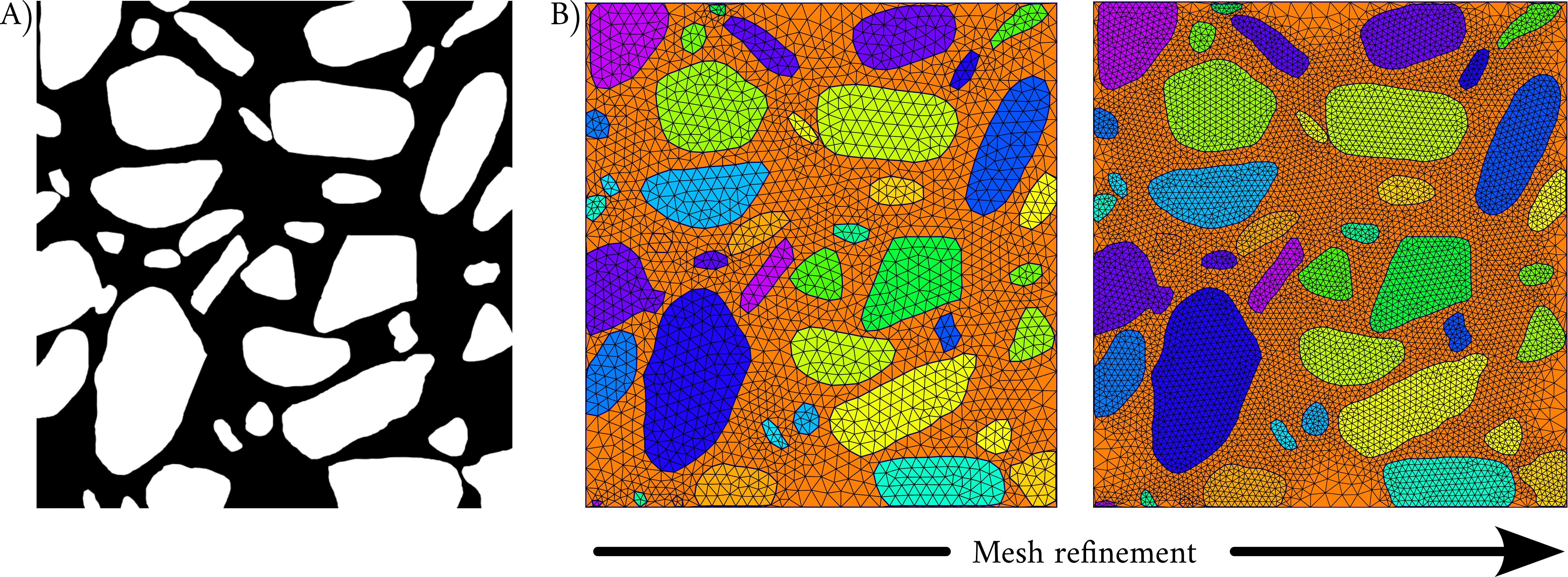}
    \caption{Remeshing of 2D concrete microstructure \cite{Coenen2021} via the elliptic decomposition \cite{Shaqfa2024_SOH}. (A) Input microstructure. (B) FEM mesh resulting from uniform isotropic remeshing of the aggregate contours achieved using a maximum of $2^5$ segments (middle inset) and $2^6$ (right inset) for reconstructing the largest aggregate contour with $N_{max} = 30$.}
    \label{fig:concrete_remeshing}
\end{figure}

For 2D microstructures, we use the 2D concrete sample shown in Fig.~\ref{fig:concrete_remeshing}A; retrieved from Coenen et al. (2021) \cite{Coenen2021}. To obtain a consistent morphology-preserving discretization of the aggregate contours, we predetermined the maximum number of segments allowed for the largest aggregate particle. We then linearly interpolated, for each particle, the maximum number of segments as a function of the contour length, with a minimum of five segments. The FEM discretization in Fig.~\ref{fig:concrete_remeshing}B corresponds to $2^5$ and $2^6$ (from left to right) maximum number of segments with $N_{max} = 30$. After remeshing the segments of the aggregate, we obtained the bulk triangulation via the Gmsh package \cite{Geuzaine2009}. The process from harmonic decomposition to remeshing all aggregate contours has taken about one minute in CPU time.

\subsubsection{Remeshing 3D stone masonry microstructure}
\par
Stone masonry microstructures are usually composed of stone inclusions and a mortar matrix that binds the stones together \cite{SHAQFA2022}. This type of microstructure is generally challenging to mesh \cite{Lu2014, Kamel2019, SHAQFA2022}. The process of generating virtual or scanned masonry microstructures begins by generating or scanning stones, followed by placing them within the wall's volume \cite{SHAQFA2022, Saloustros2023, dataset_1, Ortega2023}. This process requires generating a binding matrix that partially fills \cite{Pluijm1992} the in-between spaces of the registered stones' scans. To extract the mortar matrix, we resort to boolean operators (see Shaqfa and Beyer (2022) \cite{SHAQFA2022}), and the quality of the resulted mortar volumes depends on (i) the quality of the stones' meshes and (ii) their relative location of stones to each other, as the mortar between very closed stones can generate undesired slivers of tetrahedral elements in 3D microstructures.
\par
\begin{figure}[!ht]
    \centering
    \includegraphics[width=1.0\linewidth]{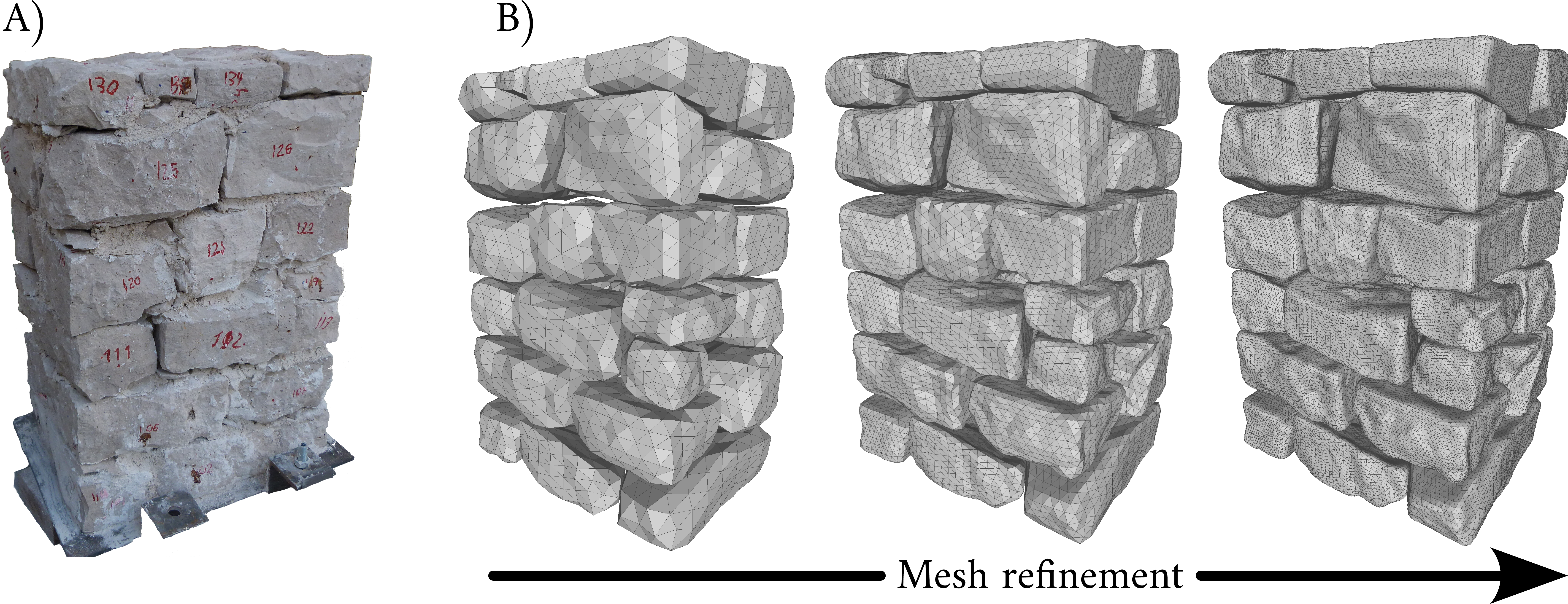}
    \caption{Isotropic remeshing of 3D stone masonry wall acquired via Structure-from-Motion (SfM). (A) A real photograph of the scanned wall \cite{Ortega2023}. (B) Uniform isotropic remeshing of the scanned stones with increasing mesh resolution from left to right.}
    \label{fig:wall_remeshing_stns}
\end{figure}

Ortega et al. \cite{Ortega2023} proposed using the Structure from Motion (SfM) imaging technique to build a digital twin (DT) of a stone masonry wall, followed by testing their mechanical behavior \cite{Ortega2024}. The stone meshes generated in their approach are not suitable for mechanical simulations. Here, we uniformly discretize one of their walls (c.f. Wall no. 1 in \cite{Ortega2023} and Fig.~\ref {fig:wall_remeshing_stns}A) with two, three, and four mesh refinement cycles of icosahedral; see Fig.~\ref{fig:wall_remeshing_stns}B (from left to right). This ensures that the mesh of all stones has an identical number of vertices and faces. Using a single-stage solving scheme, the average time spent per isotropic meshing of a stone is about: $0.647$ seconds for two refinement cycles with $I_{max} = 20$, $2.673$ seconds for three refinement cycles $I_{max} = 30$, and $9.545$ seconds for $I_{max} = 40$. Due to the surface area normalization step in Algorithm \ref{algo:iso}, we calibrated the number of iterations and time steps for one stone and generalized the parameter-setting to the rest of the wall.
\par
\begin{figure}[!ht]
    \centering
    \includegraphics[width=1.0\linewidth]{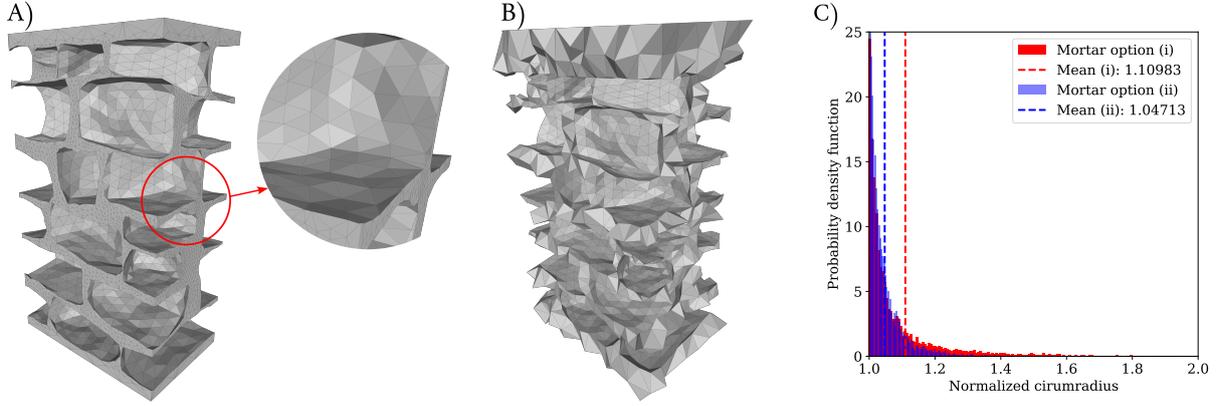}
    \caption{Surface meshes of the mortar layers resulted from different Boolean options. (A) Boundary-intersecting mortar (i) that corresponds to straight boundaries intersecting solid inclusions. (B) Boundary-conformal mortar matrix. (C) Comparison of the exterior surface quality of both mortar surfaces, where the circumradii of the second option are superior to the first one.}
    \label{fig:mortar_qlty}
\end{figure}

It is essential for numerical simulations to extract a conformal mortar-to-stone binding volume \cite{SHAQFA2022}, as such conformal meshes prevent volume overlap between the contacting regions of the stones and the mortar. To avoid confusion between the term \textit{conformal mapping} and \textit{conformal meshes}, conformal meshes refer to matching discretizations at interfaces where two material media meet. Based on how the exterior boundary of the binding volume interacts with solid inclusions, we can further classify binding matrices into: (i) an intersecting boundary, where particles cut the mortar boundary, and (ii) a nonintersecting boundary, where the matrix surface remains free of particle intersections. The first type is challenging as the remeshed stone faces will be further split along the boundary intersection points of the mortar (see case Fig.~\ref{fig:mortar_qlty}A). This split in faces can result in badly shaped triangular faces that affect the overall quality of the 3D solid mesh.
\par
To avoid the poor-quality meshes in Fig.~\ref{fig:mortar_qlty}A and obtain a boundary-conformal mortar mesh, we first assume that the mortar volume completely embeds the solid inclusions inside it. Then, we use the traditional conformal 3D FEM mesh (tetrahedra) of the mortar volume. Lastly, to obtain a boundary-conformal mortar, we delete tetrahedral elements up to the desired depth from all sides. The resulting mortar boundary is extracted and is perfectly conformal to the stones without splitting any surfaces (see Fig.~\ref{fig:mortar_qlty}B). To compare the surface qualities between the traditional mortar extraction approach and the one proposed, we computed the circumradii of the surface meshes as shown in Fig.~\ref{fig:mortar_qlty}C. It is clear that the new boundary-conformal mortar has better mesh quality and is closer to a perfect triangulation (i.e., equilateral elements with $\hat{\rho} \to 1$).

\section{Conclusion}
\label{sec:conc}
\par
Utilizing recent harmonic decomposition methods, we proposed a new morphology-preserving discretization approach to produce high-quality triangulated meshes suitable for numerical simulations. In this paper, we remesh genus-0 closed and open parametric surfaces starting from their spheroidal and hemispheroidal decomposition domains, respectively. Using the heat diffusion analogy, we resampled the spheroidal coordinates by diffusing the coordinates of large triangulated faces to enlarge small ones. We proposed and tested two main schemes for the diffusion problem: (i) an isotropic approach that merely equalizes the areas of triangular elements without considering the angular structure of the triangles. In the case of large surface protrusions, the isotropic approach can produce undesired obtuse triangles. To alleviate this effect, we also proposed a complementary (ii) anisotropic Laplace-Beltrami operator that can balance between area equalization and angular distortion of triangular faces. The mesh convergence of the anisotropic operator is generally slower than the isotropic one, and compromises the area equalization of triangulation. To monitor the quality of the surfaces, we relied on two main metrics, the distribution of triangular face areas and the normalized circumradius of triangles.
\par
We tested the diffusion approach on multiple genus-0 closed and open benchmark surfaces using multiple mesh refinements and reconstruction degrees. Based on the reconstruction degree and the corresponding computational times, we proposed a hierarchical diffusion scheme that enables efficient convergence and shorter computational times. For engineering applications, we remeshed the 2D microstructure of a concrete sample, followed by remeshing the 3D microstructure of a real stone masonry wall as representative examples of inclusion microstructures. Overall, the results revealed substantial improvements in the prescribed quality metrics, which improve the accuracy of the numerical simulations of real complex microstructures.
\par
The proposed diffusion-based approach can be used for applications beyond the main objectives of this study. For instance, we can use it to equidistantly distribute Gauss points for integrating physical quantities on 1- and 2-manifolds. Furthermore, this approach can be extended to include sampling and meshing point clouds, manifold harmonics for arbitrary surfaces with arbitrary genus, reparameterizing computer-aided surfaces (CAD) represented by Non-Uniform Rational B-Splines (NURBS), and adaptive remeshing using implicit distance functions.

\section*{Reproducibility}
\par
To test and reproduce the presented approach and results in this work, we made all the Python 3.8-compatible code openly available under the GNU license on our online repository:
\begin{itemize}
    \item GitHub: \url{https://github.com/msshaqfa/harmonic_remeshing}
\end{itemize}

\section*{Acknowledgments}
\par
The author wants to thank the Swiss National Foundation (SNSF) for partially funding this work under projects \href{https://data.snf.ch/grants/grant/211088}{P500PT\_211088} and \href{https://data.snf.ch/grants/grant/235509}{P5R5-2\_235509}. The author would like to thank Prof. Hamid Karbasian (Southern Methodist University), Prof. Ketson R M dos Santos (University of Minnesota Twin Cities), and Dr. Maja Baniček (University of Zagreb) for their constructive comments on this work.

\end{document}